\documentclass[prb,twocolumn,showpacs,preprintnumbers,amsmath,amssymb,superscriptaddress,aps,10pt]{revtex4-1}

\usepackage{graphicx}
\usepackage{dcolumn}
\usepackage{bm}
\usepackage{natbib}
\usepackage{amsmath}
\usepackage{color}
\usepackage{ulem}

\begin{document}

\title{Imaging backscattering through impurity-induced antidots \\ in quantum Hall constrictions}

\author{Nicola Paradiso}
\affiliation{NEST, Istituto Nanoscienze-CNR and Scuola Normale Superiore, Pisa, Italy}

\author{Stefan Heun}\email{stefan.heun@nano.cnr.it}
\affiliation{NEST, Istituto Nanoscienze-CNR and Scuola Normale Superiore, Pisa, Italy}

\author{Stefano Roddaro}\email{s.roddaro@sns.it}
\affiliation{NEST, Istituto Nanoscienze-CNR and Scuola Normale Superiore, Pisa, Italy}
\affiliation{Istituto Officina dei Materiali CNR, Laboratorio TASC, Basovizza (TS), Italy}

\author{Giorgio Biasiol}
\affiliation{Istituto Officina dei Materiali CNR, Laboratorio TASC, Basovizza (TS), Italy}

\author{Lucia Sorba}
\affiliation{NEST, Istituto Nanoscienze-CNR and Scuola Normale Superiore, Pisa, Italy}

\author{Davide Venturelli}
\affiliation{NEST, Istituto Nanoscienze-CNR and Scuola Normale Superiore, Pisa, Italy}

\author{Fabio Taddei}
\affiliation{NEST, Istituto Nanoscienze-CNR and Scuola Normale Superiore, Pisa, Italy}

\author{Vittorio Giovannetti}
\affiliation{NEST, Istituto Nanoscienze-CNR and Scuola Normale Superiore, Pisa, Italy}

\author{Fabio Beltram}
\affiliation{NEST, Istituto Nanoscienze-CNR and Scuola Normale Superiore, Pisa, Italy}

\date{\today}

\begin{abstract}
We exploit the biased tip of a scanning gate microscope (SGM) to induce a controlled backscattering between counter-propagating edge channels in a wide constriction in the quantum Hall regime. We compare our detailed conductance maps with a numerical percolation model and demonstrate that conductance fluctuations observed in these devices as a function of the gate voltage originate from backscattering events mediated by localized states pinned by potential fluctuations. Our imaging technique allows us to identify the necessary conditions for the activation of these backscattering processes and also to reconstruct the constriction confinement potential profile and the underlying disorder.
\end{abstract}

\pacs{73.43.-f, 72.10.Fk}

\maketitle

\section{Introduction}

The ability of scanning probe microscopy (SPM) techniques to probe and manipulate electronic states on a scale smaller than typical coherence lengths has allowed to directly visualize electron interference, the main fingerprint of quantum phenomena.  The results can be spectacular, as witnessed by the coherent electron flow pictures observed in earlier scanning gate microscopy (SGM) measurements by the Westervelt group in Harvard.\cite{Topinka2000,Topinka2001,Leroy2005}  In these experiments, the negatively biased tip of an atomic force microscope (AFM) was exploited to backscatter electrons transmitted across a quantum point contact (QPC) in a two-dimensional electron gas (2DEG). The interference among different backscattering paths produces a modulation of the transmission probability which is visible in the SGM images as fringe structures with spacing $\lambda_F/2\approx 20$~nm, where $\lambda_F$ is the Fermi wavelength. The high resolution required to resolve such structures  is \textit{not limited by the width of the tip-induced potential}: it just depends on the accuracy of tip positioning. This idea is at the basis of most experiments exploiting the SGM technique. 

In this paper we apply the same technique to study the backscattering mechanisms in wide constrictions defined in quantum Hall (QH) systems. When the bulk filling factor $\nu_b$ is integer, the current is carried by chiral edge channels.\cite{Halperin1982} Our constriction is sufficiently wide to allow full transmission when the SGM tip is far away from the constriction axis. When the negatively biased  tip is moved towards the constriction, backscattering is induced between counter-propagating edge-channels. The resulting SGM images, obtained by mapping the transmitted linear conductance  $G_T$ as a function of tip position, show that $G_T$  monotonically decreases when the tip approaches the constriction, with plateaus in correspondence to multiples of $G_0= e^2/h$. This behaviour is well known for short QPCs.\cite{Paradiso2010} However, our SGM scans over wide constrictions show a surprising fine structure consisting in small spatial oscillations of $G_T$ that lead to arc features in the SGM maps. Based on a detailed comparison with a numerical model, the emergence of such arc features can be attributed to backscattering through the constriction mediated by naturally-occurring antidot (AD) structures originating from potential fluctuations present in the constriction.

Arc structures have been observed before in SGM experiments performed on a variety of different systems, with or without magnetic field (e.g.~carbon nanotubes,\cite{Woodside2002} quantum rings,\cite{Hackens2006,Hackens2010} InAs nanowires,\cite{Bleszynski2007} and quantum dots\cite{Fallahi2005,Gildemeister2007}) and can be caused by a tip-induced modulation of either quantum interference \cite{Hackens2006} or Coulomb blockade (CB).\cite{Woodside2002,Bleszynski2007,Fallahi2005,Gildemeister2007}  Indeed, both mechanisms can be expected to play a role in artificial AD devices, as highlighted by a number of investigations in past years reviewed in Ref.~\onlinecite{Ford} and by recent experiments specifically designed to address this issue.\cite{Kou,Markus2} In our experiment, the SGM tip modulates the transport through random localized states, inducing a suitable coupling between them and the extended edge modes. For specific tip configurations which will be discussed in the paper, we induce in a highly controlled manner AD backscattering paths which would appear as conductance fluctuations in transport experiments upon pinching-off the constriction. This mechanism is absent or negligible in other tip modulation schemes such as, for instance, remote gating in quantum dots.\cite{Woodside2002,Bleszynski2007,Fallahi2005,Gildemeister2007}  It should be noted that similar arc structures have also been observed by the scanning charge accumulation technique in the QH regime.\cite{Finkelstein2000,Ashoori1}  Even if these results are linked to edge-edge scattering mediated by localized states, the mechanism behind the arc formation is different: in Refs.~\onlinecite{Finkelstein2000,Ashoori1} arcs have an angular extension which reflects the incompressible stripe modulation given by the combined action of the tip electric field and the local average field at the active edge-edge scattering center; in our work the arc angular extension rather reflects the position of the antidot with respect to the constriction borders, as discussed in the following sections.

In Ref.~\onlinecite{Paradiso2011} we studied the nature and the impact of impurity potentials on the co-propagating edge channel mixing. In our constrictions we found typically a few strong scattering centers per square micron. In order to study the behaviour of these centers as ADs linking counter-propagating edge channels crossing the constriction, we designed a constriction which contains several scattering centers between the channels. For this reason, the constriction area ($\approx$ 3~$\mu$m$^2$) in the present experiment is much larger than in our earlier work on short QPCs,\cite{Paradiso2010} which aimed at revealing the inner edge structure. 

\section{Experimental details}

The samples for this study were fabricated starting from an Al$_{0.33}$Ga$_{0.67}$As/GaAs heterostructure with a two-dimensional
electron gas (2DEG), which is confined 55~nm underneath the surface. The electron sheet density and mobility at low temperature are $n = 3.12\times 10^{15}$~m$^{-2}$ and $\mu = 4.2\times 10^{2}$~m$^2$/Vs, respectively, as determined by Shubnikov-de Haas measurements. The Hall bar was patterned via optical lithography and wet etching. Source and drain contacts were fabricated at the ends of the Hall bar by evaporation and thermal annealing of a standard Ni/AuGe/Ni/Au multilayer (10/200/10/100~nm). In order to define a constriction in the 2DEG, we fabricated two split-gates via electron beam lithography. They consist of a Ti/Au bilayer (10/20~nm). The nominal gap between the gates is 1.2~$\mu$m wide and 2.5~$\mu$m long.

Our measurements were performed with the 2DEG at bulk filling factor $\nu_b=2$ ($B= 6.450$~T). The split--gate bias was set to $-0.350$~V, so that the filling factor under the gates is $g=0$, and both edge channels of the $\nu_b=2$ bulk phase are 
sent into the constriction. The width of the constriction has been chosen large enough to allow full transmission of both edge channels, so that the source-drain transmitted  conductance, in the absence of the tip, is $G_T=\nu_b G_0 = 2e^2/h$. 

 The SGM was operated in a $^3$He cryostat (base temperature 300~mK). Sample temperature was 400~mK, as calibrated with a Coulomb blockade thermometer. The cryostat is equipped with a superconducting coil which provides magnetic fields of up to 9~T.
Details of our SGM setup are reported in Ref.~\onlinecite{Paradiso2010}. SGM maps are obtained by scanning a negatively biased AFM tip over the constriction area and acquiring for each position $(x_t,y_t)$ the corresponding source-drain transmitted  conductance $G_T(x_t,y_t)$.

\section{Experimental results}

\begin{figure}[tb]
\includegraphics[width=\columnwidth]{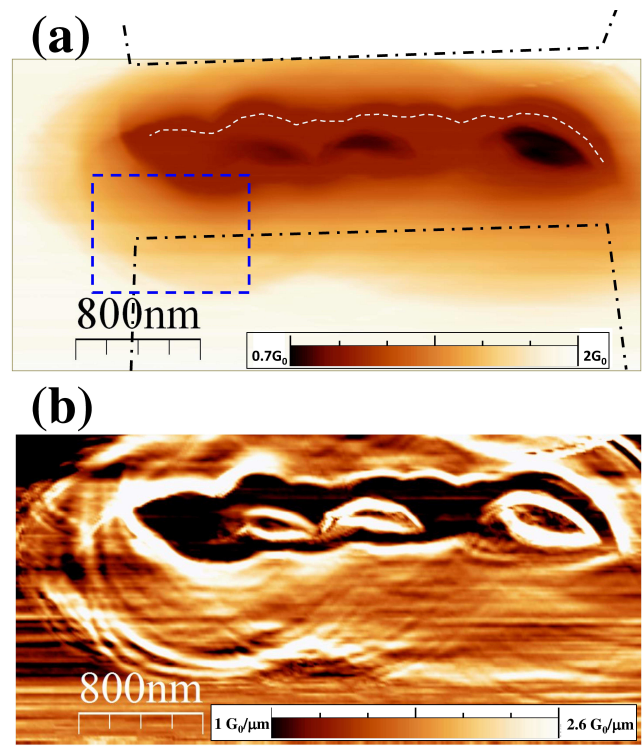}
\caption{(a) SGM scan over a 1.2~$\mu$m wide and 2.5~$\mu$m long constriction defined in a $\nu_b=2$ quantum Hall system ($B=6.450$~T, $V_{tip}=-6.0$~V). The map displays the transmitted conductance $G_T(x_t,y_t)$ plotted as a function of the tip position. As a reference, we indicated the middle of the $G_T=G_0$ plateau with a dashed white line. (b) Gradient map $|\mathbf{\nabla}G_T(x_t,y_t)|$ which emphasizes short scale modulations of the $G_T$ signal. A fine pattern of arc structures is observed, indicating the occurrence of peculiar backscattering processes.}
\label{fig:general}
\end{figure}

Figure~\ref{fig:general}(a) shows the result of a scan over the whole constriction area, performed at $B=6.450$~T and $V_{tip}=-6.0$~V. When the SGM tip is far from the constriction, there is no backscattering between counter-propagating edge states, hence the source-drain conductance transmitted across the constriction is $G_T = 2 G_0 = 2e^2/h$.  When the tip approaches the constriction axis, tunneling between the counter-propagating edges is induced, and $G_T$ is reduced, similarly to what was observed in previous SGM experiments on short QPCs.\cite{Aoki2005,Paradiso2010} The detailed shape of contour lines at equal conductance is different from device to device (for six devices measured in our experiments) and depends on the actual disorder potential inside the constriction. An extended area with constant conductance belongs to the $G_T=G_0$ plateau and is indicated in Fig.~\ref{fig:general}(a) with a dashed white line.

The present QPC design allows to further investigate the backscattering mechanism which governs transport in the constriction. A number of interesting experimental features can be highlighted by plotting the $|\mathbf{\nabla}G_T(x_t,y_t)|$ map (shown in Fig.~\ref{fig:general}(b)), which allows to emphasize short-range variations of $G_T$. Clear resonance patterns can be observed along a set of arc-shaped regions. The arcs are regular, and their centers typically lie in the $G_T=0$ region, i.e.~close to the center of the constriction. Interestingly, the arc structures are completely suppressed within the $G_T=G_0$ plateau.

\begin{figure}[tb]
\includegraphics[width=0.8\columnwidth]{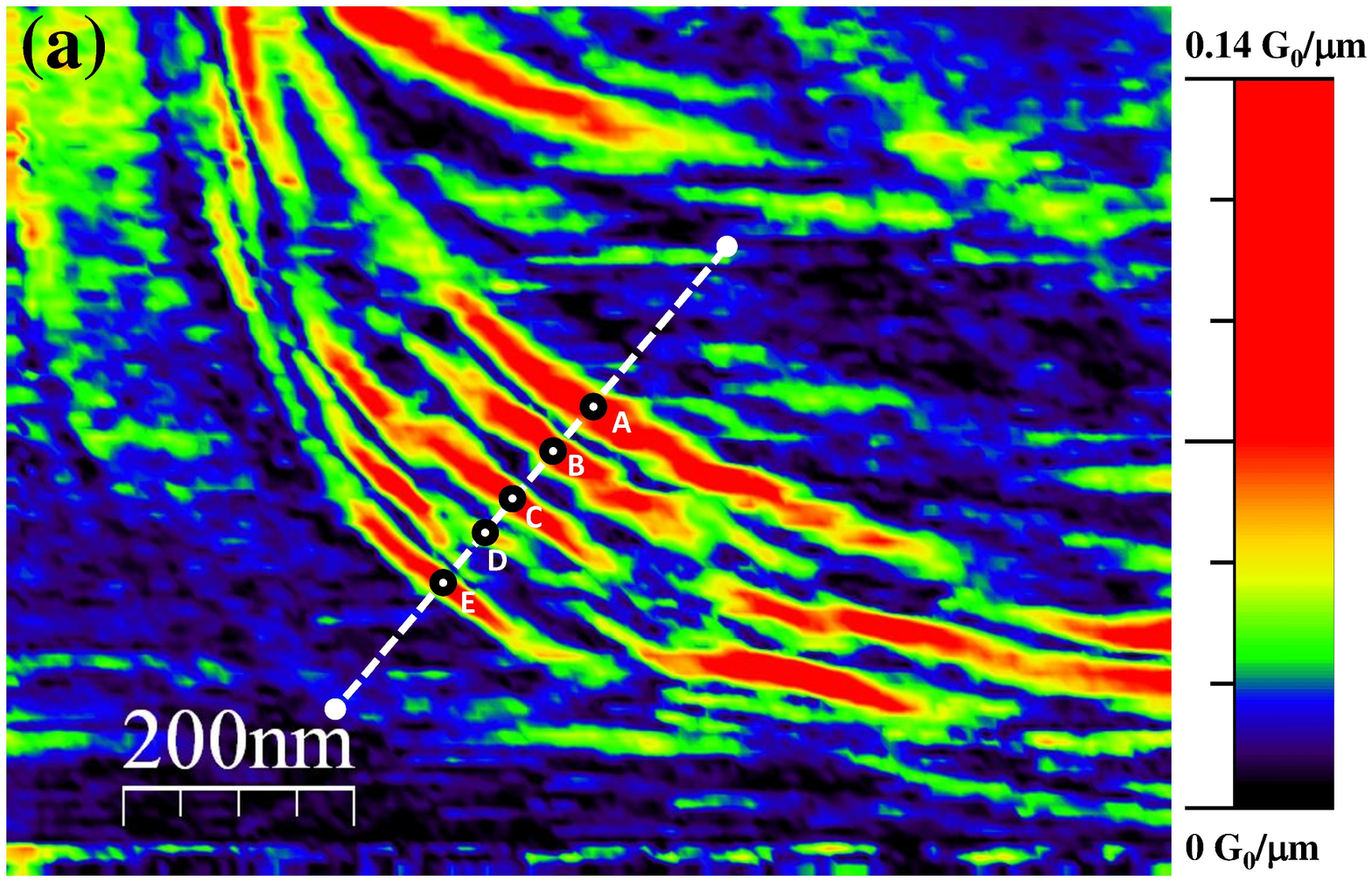}
\includegraphics[width=0.8\columnwidth]{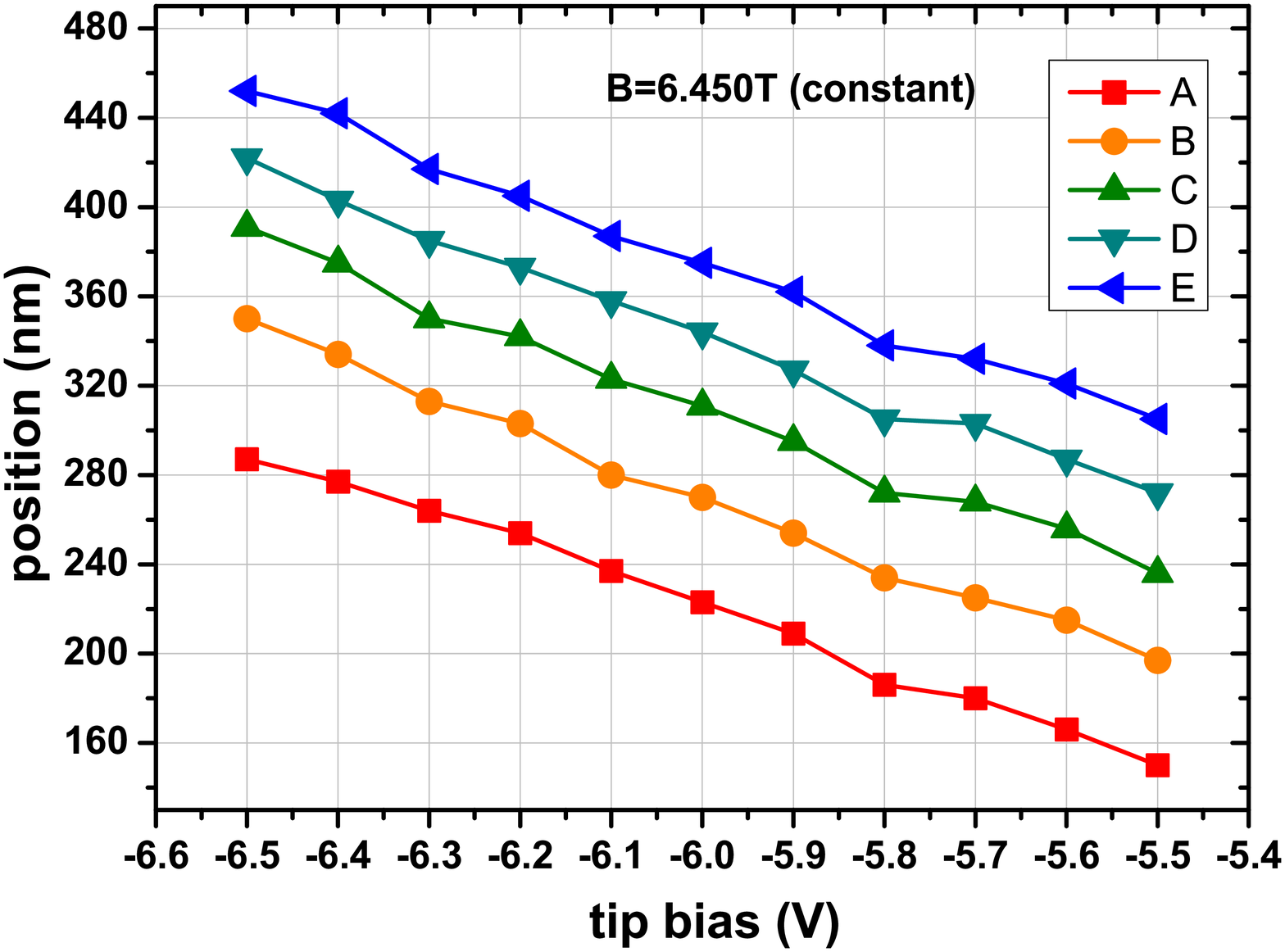}
\includegraphics[width=0.8\columnwidth]{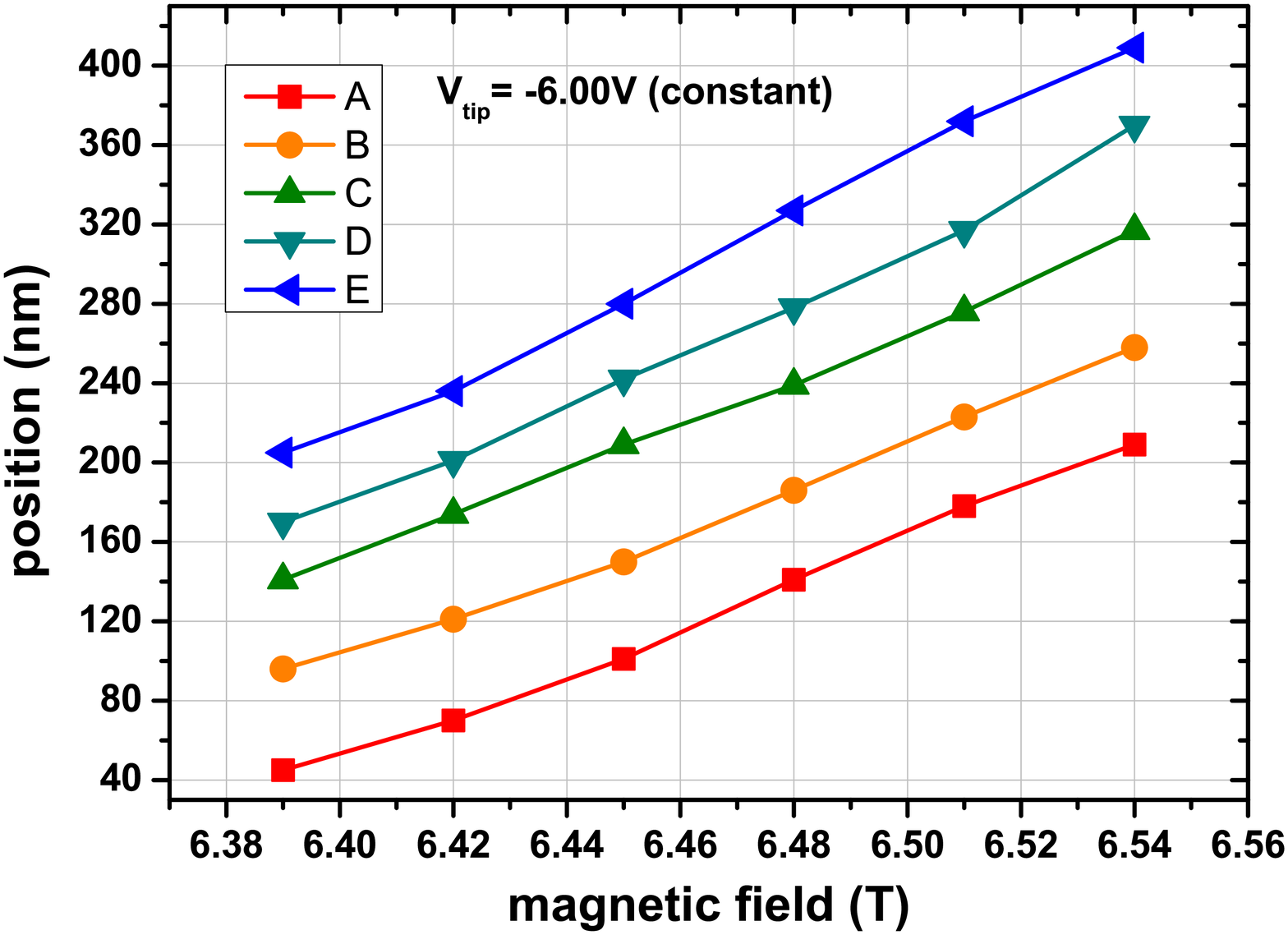}
\caption{(a) High resolution scan over the region indicated by the dashed blue box in Fig.~\ref{fig:general}(a). $B=6.510$~T and $V_{tip}=-6.0$~V. Measuring a series of such images allows to follow the position of five arcs, labeled  A--E, when either tip voltage $V_{tip}$ or magnetic field $B$ are varied.
(b) Displacements of the A--E peaks as a function of $V_{tip}$ at constant $B= 6.450$~T. Displacements are linear in $V_{tip}$, with approximately the same slope for all five structures, so that their spacing is nearly constant.
(c) The corresponding arc displacements when $B$ is varied, at constant $V_{tip}=-6.0$~V. Also in this case, an approximately  linear dependence is observed.}
\label{fig:zoom}
\end{figure}

Transport data in one of the resonance regions is analyzed in finer details in Fig.~\ref{fig:zoom}(a), reporting a scan ($B=6.510$~T and $V_{tip}=-6.0$~V) over the region corresponding to the dashed blue rectangle in Fig.~\ref{fig:general}(a).
We study the evolution of the resonances as a function of applied magnetic field $B$ and tip voltage $V_{tip}$. A careful analysis of these images shows that both an increasing magnetic field and an increasingly more depleting (i.e.~more negative) scanning tip induce a shift of the resonances towards the outer side of the arcs. The effect can be appreciated in Figs.~\ref{fig:zoom}(b) and (c), where the position of the maxima of the conductance gradient 
modulus ($|\mathbf{\nabla} G_T(x,y)|$) along the section indicated in Fig.~\ref{fig:zoom}(a) is reported. The distance between two adjacent maxima in $|\mathbf{\nabla} G_T(x,y)|$ corresponds to \textit{half} an oscillation period in $G_T(x,y)$. We can estimate from the roughly linear curves in Fig.~\ref{fig:zoom}(b,c) the typical periodicity of conductance oscillations in the radial space direction ($\lambda_R\simeq 80\pm20$~nm, roughly independent of $B$ and $V_{tip}$), as well as in the magnetic field ($\lambda_B\simeq 73\pm18$~mT) and tip bias ($\lambda_V\simeq 0.64\pm0.16$ V) for fixed tip position. As we shall argue in  Sec.\ref{sec:origin}, oscillations over these scales provide evidence for the presence of backscattering events mediated by localized AD channel states. 

\section{Discussion}

\begin{figure}[!ht]
\includegraphics[width=\columnwidth]{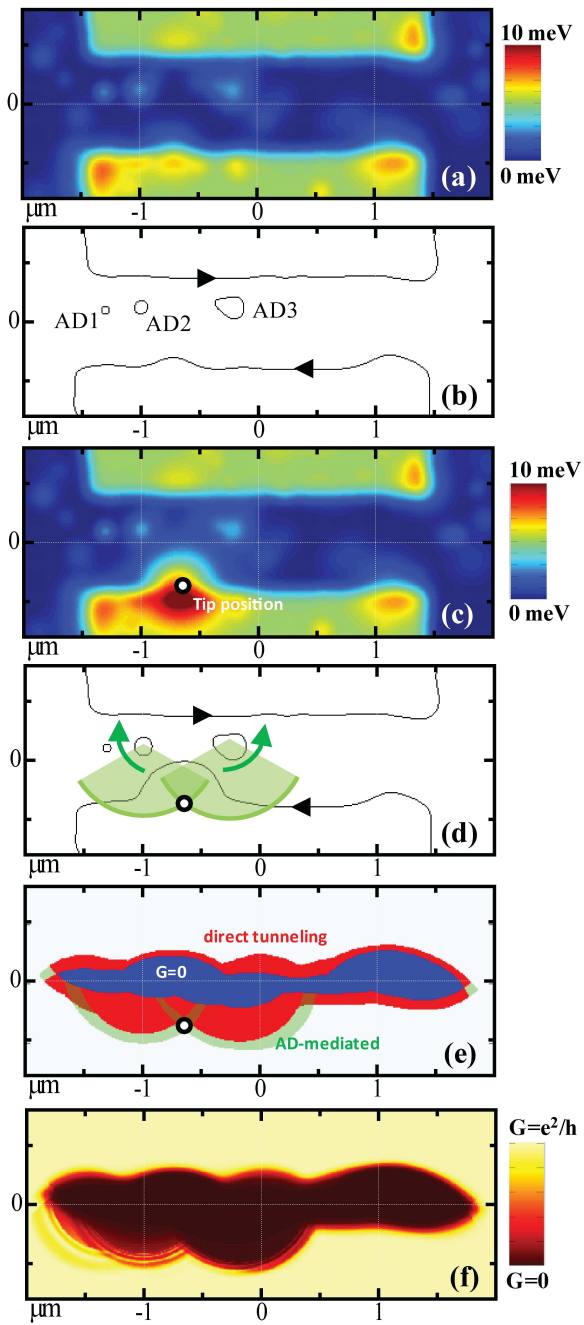}
\caption{(a) Artificial constriction landscape with disorder and (b) the corresponding edge structure. (c,d) Modified constriction after the addition of a tip depletion: the tip displaces the edge trajectory and activates two backscattering channels through AD2 and AD3. (e) Direct and antidot-mediated regions in the scan and (f) heuristic calculation of the $G_T(x_t,y_t)$ map.}
\label{fig:model1}
\end{figure}

In order to analyze the mechanism behind the tip-induced backscattering, we have performed a numerical calculation of $G_T$ as a function of the tip position $(x_t,y_t)$ on the basis of a realistic potential landscape $U(x,y)$ (defined on a lattice with $10\,{\rm~nm}$ stepsize) which models a single-edge constriction and the disorder present in the sample (we neglect the spin degree of freedom). While a specific potential landscape is considered in the following, qualitatively similar results could be obtained regardless of the precise form of $U(x,y)$. The constriction is obtained by imposing a potential step of $25\,{\rm meV}$, sufficient to strongly deplete the 2DES. The electron liquid is characterized by a Landau level gap $\Delta\approx 10\,{\rm meV}$, of the order of the cyclotron gap in our experiment $\hbar\omega_c=11.1\,{\rm meV}$.
The 25~meV potential step is imposed under the gate position plus a smooth depletion corona of $\approx 200$~nm, while impurities are introduced by adding a set of random Gaussian bumps. A specific realization $U(x,y)$ is reported in Fig.~\ref{fig:model1}(a), with three prominent potential hills (light blue) on the center-left side of the constriction which is defined by the top and bottom gate barriers (orange-red). The guiding center (trajectory) of the first QH edge channel in the constriction is obtained as the equipotential line satisfying $U(x,y)=\Delta/2$, where we take the constriction saddle point as the energy origin. Such an edge-extraction procedure is expected to hold exactly in the $B\to\infty$ limit and to be reasonable as long as the relevant edge features are not much smaller than the magnetic length $\ell=\sqrt{\hbar/eB}$ ($\approx10\,{\rm~nm}$ in our experimental configuration). The edge channel trajectory for the potential of Fig.~\ref{fig:model1}(a) is presented in Fig.~\ref{fig:model1}(b) and shows the formation of three localized edge states around the locally-depleted regions denoted by AD1, AD2, and AD3.

The influence of the SGM tip is introduced in the calculations as an additional moveable depletion spot located at the tip position $(x_t,y_t)$ and described by the potential
\begin{equation}
U_{tip}(x,y)=A_t\exp\left[-\Delta r^2/2R_t^2\right] \; ,
\label{eq:potential}
\end{equation}
where $\Delta r^2=(x-x_t)^2+(y-y_t)^2$, $R_t=200\,{\rm~nm}$, and $A_t=$25~meV, i.e.~inducing complete depletion under the tip. It is worth to remark that in the experiment we rather observe the effect of the tail of the Gaussian potential of Eq.~(\ref{eq:potential}). Therefore the energy dispersion and the electron density slope at the edge of the constriction are considerably smoother when the tip is present. The resulting potential landscape for $x_t=-650\,{\rm~nm}$ and $y_t=-300\,{\rm~nm}$ is shown in Fig.~\ref{fig:model1}(c) and the corresponding edge channel trajectories in Fig.~\ref{fig:model1}(d). As one could naively expect, the tip is able to deviate the edge trajectory and induce new backscattering channels between the top edge (TE) and bottom edge (BE) of the constriction. The determination of the backscattering amplitude for a given edge configuration is highly non-trivial and will depend on a host of (often unknown) experimental parameters, including edge chirality, drift velocity, dephasing times, edge potentials and energy distributions, etc. Despite this, the key features of the conductance scan $G_T(x,y)$ can be captured by taking into account the {\it direct} backscattering BE $\to$ TE as well as the {\it indirect} backscattering involving a {\it single} antidot, for instance BE $\to$ AD2 $\to$ TE. For sake of simplicity and because of their expected smaller and less likely contribution, we instead neglected backscattering channels involving conduction through multiple AD in series.

As the simplest possible model we assume that an appreciable backscattering takes only place if the minimum distance between two edge trajectories is less than a cut-off interaction length of 200~nm. Above this limit the two edges will be considered as non--interacting. In Fig.~\ref{fig:model1}(e) we show how this criterion, combined with the calculated guiding centers of the edge channels described above, allows us to determine the regions where $G_T=0$ (in blue, when the two edges are fully backscattered), those where $G_T=\nu_be^2/h$ (in white, when no interaction is present) and the regions where a \textit{direct} (red) or \textit{AD-mediated} (green shade) backscattering is taking place. The presence of arcs in the conductance map can now be accounted for by assuming that each AD-mediated backscattering path gives rise to conductance oscillations arising from either coherent interference between multiple trajectories carrying different Aharonov-Bohm (AB) phases or charging effects. Both mechanisms can give rise to oscillations, as discussed in Sec.~\ref{sec:origin}.  The resulting plot is shown in   Fig.~\ref{fig:model1}(f) -- see Appendix for details. By comparing it with Fig.~\ref{fig:general}(b) one realizes that our basic model reproduces the general features, including the arc structures, of the SGM scan. In this respect it is important to note that while the chosen phenomenological cut-off length of 200~nm can be considered to be arbitrary, the general features of the simulations are largely independent of the precise value of this parameter.

In particular, the simulation shows that the centers of the arcs observed in the simulated SGM maps are approximately located on an AD, as indicated by the overlays in Fig.~\ref{fig:model1}(d) and in the shaded regions in Fig.~\ref{fig:model1}(e). This proves the origin of the resonances and allows us to map the location of the most relevant scattering centers in our constriction potential landscape. As clarified by Fig.~\ref{fig:model1}(d), when moving along these arcs the tip mediates a suitable interaction between the TE and the BE channels through the AD, thus activating the backscattering mechanism. Different tip positions along the arc correspond to similar values of AD-mediated edge channels coupling. On the contrary, moving the tip towards or away from the AD will modify the coupling and also the area of the AD because of the long range depletion due to the tail of the tip potential. This can also be seen from a comparison between Fig.~\ref{fig:model1}(b) and Fig.~\ref{fig:model1}(d): when the tip is moved towards an AD, for example AD3, the AD area increases.

\section{Interpretation of the conductance oscillations}\label{sec:origin}

In this section we focus on the origin of the observed resonances. Backscattering paths mediated by a localized electronic state are generally expected to give rise to periodic oscillations as a function of the magnetic field as well as of tip position and bias. As argued in the following, this kind of transport features can originate from two different phenomena: coherent interference between different paths encircling, for instance, an isolated antidot (AB effect); and/or CB oscillations due to charging effects. The exact prevalence of one or the other regime can be subtle, and different working conditions have been identified and studied in recent experiments depending on sample details.\cite{Kou,Markus2} Our experimental data do not allow to conclusively rule out one of the two possibilities, therefore in the following we propose a basic model which considers both effects.

The AB effect can be accounted for by assuming that an AD is induced by a relatively smooth potential fluctuation, and that the tunneling region between the AD and an edge channel could be locally approximated by a parabolic saddle-point potential.\cite{TunnelingJain} In this case the conductance $G_T$ due to backscattering through the AD can be written as:\cite{TunnelingKivelson}
\begin{equation}
\Delta G_T=\frac{e^2}{h}{\left| \frac{t_{1}t_{2}}{1-r_{1}r_{2}\exp({i\varphi})} \right|}^2 \; ,
\label{eq:DGt}
\end{equation}
\noindent where $t_1$ and $t_2$ are the transmission amplitudes between BE and AD, and between AD and TE, respectively, while 
(under the assumption of a symmetric tunneling region) $r_j=i\sqrt{1-|t_j|^2}$ ($j=1,2$) are the relative reflection amplitudes.\cite{Zeilinger} Here $\varphi=2\pi\Phi/\Phi_0$ is the AB phase corresponding to the magnetic flux $\Phi=B\Omega$ enclosed within the effective area $\Omega$ of the AD, $\Phi_0=h/e\approx4.13\,{\rm mT \cdot\mu m^2}$ being the flux quantum, giving rise to oscillations with a period $\Delta B = \Phi_0/\Omega$. 

Coulomb interaction can influence this simple picture in a number of different ways, in particular due to the emergence of compressible stripes and to the importance of CB phenomena in the transport. Charging effects in the localized backscattering center can be taken into account (see Refs.~\onlinecite{Sim,Kataoka,Ford,Ihn,HoutenBenakkerStaring}) resulting in a periodicity of the Aharonov-Bohm oscillations which is rescaled with a factor $\alpha$, i.e.~$\Delta{B}\rightarrow \Delta{B}/\alpha$. Taking $\alpha$ as the number of edge states around the AD (see Refs.~\onlinecite{FordSimp,Goldman,KataokaFord,Kou}), in our case we have $\alpha=2$. The observed periodicity of $73\pm18$ mT would thus correspond, in the case of pure AB oscillations, to an area of $0.057\pm0.014\,{\rm \mu m^2}$ or a circle with radius $R\approx 140$~nm. Considering also charging effects, we obtain an area of $0.113\pm0.028\,{\rm \mu m^2}$ or a circle with radius $R\approx 190$~nm.

Important information about the backscattering center can also be gained by comparing the field, tip position, and tip bias dependence of the oscillations. For instance, looking at data reported in Fig.~\ref{fig:zoom}, one can deduce that an increase of $B$ by $\lambda_B$ can be compensated by moving the tip away from the AD by a radial distance $\lambda_R$. This allows to determine how the tip radial distance $x$ influences the effective area of the scatterer ($d\Omega/dx = -\Phi_0/B\lambda_R<0$) and to conclude that the backscattering center becomes larger when the tip depletion spot is moved towards it, i.e.~that it behaves consistently with our AD interpretation.

\begin{figure}[tb]
\includegraphics[width=\columnwidth]{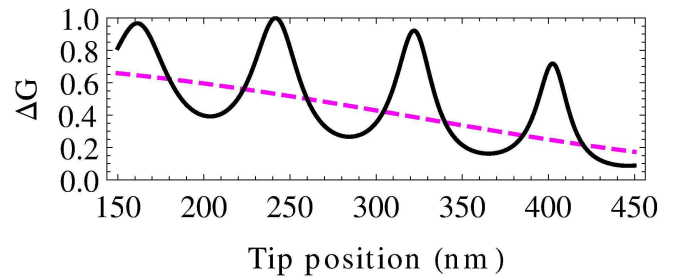}
\caption{(Color online) Conductance oscillations (solid line) and tunneling amplitudes $|t_1 t_2|^2$ (dashed line) as a function of the tip position $x$, assuming $\eta\simeq 10^{-5}$~nm$^{-2}$ and $|t_2|^2=0.75$.}
\label{fig:oscill}
\end{figure}

Figure~\ref{fig:oscill} shows a simple prediction for the AD conductance as a function of the tip position $x$ assuming \cite{TunnelingJain} a dependence of the tunneling parameter $t_1 = e^{-\eta x^2}$ with a heuristic $\eta=10^{-5}\,{\rm nm^{-2}}$ (on the side of the edge deflected by the tip potential) and a constant $|t_2|^2=0.75$ (describing the coupling to the opposite edge which is to first order unperturbed by the tip).
It should be noted that we have not taken into account any effect of temperature, dephasing, or self-averaging mechanisms, which are expected to further broaden the shape of the resonances as well as to reduce the visibility of the oscillations. This can explain the fact that only few small-amplitude oscillations are detectable in our experiment.

\section{Conclusions}

We performed a SGM mapping of the transmission through a wide constriction at filling factor $\nu_b=2$. Beyond observing the usual selective backscattering of the two spin-split edges, we consistently observe arc-shaped resonances with a typical limited extension in both the radial and angular direction. We demonstrate based on a numerical model that such transport features are consistent with AD-mediated backscattering events modulated by the tip potential tails and activated by the deflection of the edge trajectory caused by the core of the tip potential. The evolution of the resonances as a function of the SGM position, the tip potential, and the magnetic field was analyzed also taking into account effects due to the AB phase and charging effects. Our imaging technique allows to reconstruct prominent features of the confinement potential including the effective profile of the gate and the approximate position of strong scattering centers which are nucleating AD structures in the measurement.

\begin{acknowledgments}
We acknowledge financial support from the Italian Ministry of Research (MIUR-FIRB projects RBIN045MNB and RBID08B3FM). We gratefully acknowledge discussions with Rosario Fazio.
\end{acknowledgments}

\appendix*

\section{Percolation model}

The model we present here aims at reproducing the general features of the SGM scans which derive from the edge geometry as perturbed by the biased tip. The constriction conductance $G_T$ was thus calculated using two basic approximations which are expected to give a qualitatively correct estimate of the general behaviour of the edge system as a function of tip position. 

For  any couple of points  $\vec{x}_B=(x_B,y_B)$ and $\vec{x}_T=(x_T,y_T)$ located on the BE and TE  channels of Fig.~\ref{fig:model1}(c,d), respectively, we assign a probability  of  tunneling which, consistently with the definitions of Eq.~(\ref{eq:DGt}),
 is exponentially suppressed with the distance $|\vec{x}_B-\vec{x}_T|$, i.e. 
  $|t(\vec{x}_B,\vec{x}_T)|^2=e^{-|\vec{x}_B-\vec{x}_T|^2/L^2}$ 
   (here  $L=100\,{\rm nm}$ is a  phenomenological parameter describing the tunneling range). By integration over the whole extension of the channels this yields the following dimensionless quantity 
\begin{equation}
R_{BE\to TE} = 
\left[{\sum_{\vec{x}_B,\vec{x}_T}e^{-\frac{|\vec{x}_B-\vec{x}_T|^2}{L^2}}}\right]^{-1} \; ,
\label{eq:tunres}
\end{equation}
which plays the role of an effective tunneling resistance between the two edges. Assuming that the main interaction takes place in  a single localized but extended region where all the terms $t(\vec{x}_B,\vec{x}_T)$ are almost constant,  and assuming that 
coherent effects  are effectively washed out in the limit of a large number of terms $t(\vec{x}_B,\vec{x}_T)$ contributing to the percolation,  the value of the conductance $G_T$ for the $BE\to TE$ transition  
can be phenomenologically estimated as follows 
\begin{equation}
G_T\simeq \frac{e^2}{h}\frac{R_{BE\to TE}}{R_{BE\to TE}+1} \; ,
\label{eq:QPCconductance}
\end{equation}
(corrections being suppressed as the number of $t(\vec{x}_B,\vec{x}_T)$ entering in the process increases).
Notice that  Eq.~(\ref{eq:QPCconductance}) gives the correct values in the limit $R_{BE\to TE}=0$ ($G_T=0$) and $R_{BE\to TE}\to\infty$ ($G_T=e^2/h$).
Beyond direct tunneling, we also take into account backscattering events mediated by single AD structures (in the sense that series of multiple AD hopping processes are not taken into consideration)
as in the case indicated in Fig.~\ref{fig:model1}(d). To do so we include contributions of the same form as in Eq.~(\ref{eq:DGt}), where the effective tunneling amplitudes $t_1$ and $t_2$
connecting respectively BE $\to$ AD and AD $\to$ TE, 
are computed along the lines of  Eq.~(\ref{eq:QPCconductance}), i.e. 
\begin{eqnarray}
t_{1} & = & \sqrt{\frac{R_{BE\to AD}}{R_{BE\to AD}+1}} \quad\mbox{and} \\
t_{2} & = & \sqrt{\frac{R_{AD\to TE}}{R_{AD\to TE}+1}} \; .
\end{eqnarray}
The global tunneling resistance for a given tip configuration is finally calculated as an \textit{incoherent}  sum of parallel resistances coming from the direct $BE\to TE$ tunneling and all the $BE\to AD\to TE$ channels through a single AD. For instance, the specific configuration shown in Fig.~\ref{fig:model1}(d) is dominated by backscattering events involving conduction through AD2 and AD3 (the choice of summing these terms in an incoherent fashion is motivated by the large areas involved in the process). 

The final result of the numerical simulation can be seen in the colorplot of Fig.~\ref{fig:model1}(f), where one can discern many geometrical features which are remarkably similar to the ones observed in the experiment. It is important to stress that different random potential landscapes lead to qualitatively similar results in terms of $G_T(x_t,y_t)$. The shape of the pinch-off region $G_T=0$ presents round-shaped boundaries and a set of arc fringes that can be observed in the $0<G_T<e^2/h$ region. Fringes in this case signal the presence of tunneling paths mediated by an AD.
These tunneling channels are only active along arcs in the $G_T$ scan, in good agreement with features observed in the actual experimental data.
The simulation shows that the arc center is approximately located on an AD  in $U(x,y)$, as visible from the overlays in Fig.~\ref{fig:model1}(d). The SGM scan can thus be utilized to gain access to various features of the constriction potential landscape. As visible from Fig.~\ref{fig:model1}(d), along these arcs the tip potential is able to couple either the TE or the BE edge to the AD and to activate the backscattering mechanism. Different positions along the arc correspond to similar values of edge-AD coupling. On the other hand, moving towards or away from the AD modifies the tunneling amplitude and the area of the AD because of the long range depletion due to the tail of the tip potential. The radial extension of the fringes is naturally limited in space because if the tip is too close to the AD, the constriction edge will simply merge with the AD, while if the tip is too far from it, the tunneling will rapidly become exponentially suppressed.

\end{document}